\documentclass[prd,aps,floats,twocolumn,nofootinbib]{revtex4-1}
\usepackage{slashed}
\usepackage{mathtools}
\usepackage{amsfonts}
\usepackage{amssymb}
\usepackage{epsfig}
\usepackage{empheq}
\usepackage{mathrsfs}
\usepackage{xcolor}
\usepackage{hyperref}
\hypersetup{
    colorlinks=true,
    linkcolor=blue,  % Colors equation and section references
    citecolor=blue,  % Colors citation brackets/numbers
    urlcolor=blue    % Colors web links
}

\usepackage{bm}

\usepackage{appendix}

\begin{document}
\newcommand*\widefbox[1]{\fbox{\hspace{2em}#1\hspace{2em}}}
\newcommand{\m}[1]{\mathcal{#1}}
\newcommand{\nn}{\nonumber}
\newcommand{\ph}{\phantom}
\newcommand{\eps}{\epsilon}
\newcommand{\be}{\begin{equation}}
\newcommand{\ee}{\end{equation}}
\newcommand{\bea}{\begin{eqnarray}}
\newcommand{\eea}{\end{eqnarray}}
\newcommand{\cH}{{\cal H}}
\newtheorem{conj}{Conjecture}

\newcommand{\plk}{\mathfrak{h}}

%%%%%%%%%%%%%%%%%%%%%%%%%%%%%

\title{A Quasi-local Entropy for Black Hole Space-times}

\date{}

\author{Raymond Isichei}
\email{ri221@ic.ac.uk}

\affiliation{Abdus Salam Centre for Theoretical Physics, Imperial College London, Prince Consort Rd., London, SW7 2BZ, United Kingdom}

\begin{abstract}
We show that for static \& spherically symmetric black hole geometries, the angular components of the Brown-York tensor on a finite radius boundary can be interpreted as the product of the Tolman-Ehrenfest temperature and a quasi-local entropy. This entropy vanishes at spatial infinity and monotonically increases as radius decreases, taking its maximum value on the horizon where it is equal to the Bekenstein-Hawking entropy. Thus, the behaviour of this entropy provides explicit verification of the covariant entropy conjecture.
\end{abstract}

\date{28 Aug 2026}

\maketitle

%\section{Thermogravity}
\textit{\textbf{A tale of 2 perspectives:}} In contrast to all other known branches of equilibrium thermodynamics, gravitational thermodynamics is unique as the location of the thermal system boundary is of central importance. Consider the thermal system to be a manifold $M$ endowed with a metric tensor $g_{\mu \nu}$, or $(M, g_{\mu \nu})$ bounded by a boundary manifold $\partial M$ with induced metric $\gamma_{ij}$, or $(\partial M, \gamma_{ij})$. When $(\partial M, \gamma_{ij})$ is at the largest extent possible, the global free energy $F_{G}$ for $(M, g_{\mu \nu})$ is $F_{G}=E_{G}-T_{H}S$ \cite{Hawk1}, where $E_{G}$ is the global energy \cite{Hawk2}, $T_{H}$ is the global Hawking temperature of $(M, g_{\mu \nu})$ \cite{Hawk3} and $S=A_{H}/4$ is the Bekenstein-Hawking entropy where $A_{H}$ is the area of the event horizon \cite{Bek1, Bek2, Bek3}. However, when $(\partial M, \gamma_{ij})$ is at finite extent, so that parts of $(M, g_{\mu \nu})$ exist ``outside" of $(\partial M, \gamma_{ij})$, $F_{G}$ loses its validity. Instead, the relevant free energy on $(\partial M, \gamma_{ij})$ is the quasi-local free energy $F_{QL}=E_{QL}-T_{TE}S$ \cite{York}, in which $E_{QL}$ is the total gravitational energy of $(M, g_{\mu \nu})$ within $(\partial M, \gamma_{ij})$ \cite{BrownYork}, and $T_{TE}$ is the quasi-local Tolman-Ehrenfest temperature on $(\partial M, \gamma_{ij})$ \cite{TE1, Sant1, Sant2, Sant3}. As such, $F_{QL}$ contains all thermodynamic information of $(M, g_{\mu \nu})$ ``interior" to $(\partial M, \gamma_{ij})$ \cite{BrownYork2}. Yet, while $E_{QL}$ and $T_{TE}$ vary according to the location of $(\partial M, \gamma_{ij})$, the entropy in $F_{QL}$ always obeys $S=A_{H}/4$ \cite{York, BrownYork, BrownYork2}.\

This perspective appears to be irreconcilable with various entropy bounds motivated by the holographic principle which state that the entropy for a geometric volume $V$ bounded by a codimension-2 surface with area $A$ must satisfy $S\leq A/4$ \cite{Bek4, Bek5, Suss1, Suss2, thooft}. The most precise of these is Bousso's covariant entropy conjecture  which is stated as follows \cite{Bousso, Wald}: consider a space-like surface $(B, \sigma_{ab})$ in $(\partial M, \gamma_{ij})$ endowed with a null vector field $k^{a}$ which is everywhere orthogonal to $(B, \sigma_{ab})$. Let the expansion of null geodesics tangent to $k^{a}$, defined as $\theta=\nabla_{a}k^{a}$ be such that $\theta\leq 0$ on $(B, \sigma_{ab})$. Starting on $(B, \sigma_{ab})$ and following the null congruence until $\theta \rightarrow -\infty$ defines a light-sheet $L$. The entropy flux $S_{L}$ through $L$ must obey $S_{L}\leq A_{B}/4$. More generally, $L$ may begin on a space-like surface $(B, \sigma_{ab})$ and end on another space-like surface $(B', \sigma_{ab}')$ with both having $\theta\leq0$ \footnote{As is the case for ingoing radial null geodesics in spherical symmetry}. The corresponding entropy bound for this scenario is $S_{L}\leq (A_{B}-A_{B'})/4$ \cite{Wald}. Yet, according to the usual quasi-local perspective \cite{York, BrownYork, BrownYork2}, if $(B, \sigma_{ab})$ and $(B', \sigma_{ab}')$ are the boundaries of the thermal system, then $S_{L}=0$ is always true and space-time geometry has no intrinsic entropy outside of an event horizon \cite{F-1}.\

We believe this tension can be resolved when $(M,g_{\mu \nu})$ is a static, spherically symmetric black hole space-time. When $(\partial M,\gamma_{ij})$ is a finite radius boundary within $(M, g_{\mu \nu})$, we propose the existence of a \textit{quasi-local entropy} on $(\partial M,\gamma_{ij})$. The derivation of this entropy is contingent on one postulate, namely, that the thermal properties of $(\partial M, \gamma_{ij})$ are completely determined by the Brown-York tensor $\bar{\Pi}_{ij}$ on $(\partial M, \gamma_{ij})$. Given this premise, we revisit the relationship between surface gravity and temperature, while considering the ramifications of this for components of $\bar{\Pi}_{ij}$ on $(\partial M, \gamma_{ij})$. The quasi-local entropy, which vanishes asymptotically and is equal to the Bekenstein-Hawking entropy on an event horizon follows as a natural consequence of this analysis. For this analysis we will use natural units in which $\hbar=c=G=1$. \

\textit{\textbf{Formalism}}: Given any $d$-dimensional static Euclidean black hole space-time, $(M, g_{\mu \nu})$, we consider a $(d-1)$-dimensional space-like boundary $(\partial M, \gamma_{ij})$ containing the euclidean time coordinate $\tau$, with $0\leq \tau \leq \beta_{H}$ where $\beta_{H}$ is the global Hawking inverse temperature of $(M, g_{\mu \nu})$. In spherical symmetry, we may further consider a $(d-2)$-dimensional surface $(B, \sigma_{ab})$ which is topologically an $S^{d-2}$ sphere. Within the context of Euclidean gravitational thermodynamics, these geometric definitions can be used to define the equilibrium free energy $-\beta F=-I^{E}$, where
\begin{equation}
    -I^{E}=\frac{1}{16\pi }\int_{M}d^{d}x\sqrt{g}\ \mathcal{L}_{M}+\frac{1}{8\pi }\int_{\partial M}d^{d-1}x\sqrt{\gamma} \ \mathcal{L}_{\partial M}
    \label{I1}
\end{equation}
is the total gravitational action evaluated on solutions to the classical equations of motion for $\mathcal{L}_{M}:=\mathcal{L}_{M}[g_{\mu\nu}, R^{\mu}_{\nu \lambda \tau}]$ \cite{Hawk1}. On $(\partial M, \gamma_{ij})$ with outward (inward) facing normal $n=\pm n^{\mu}\partial_{\mu}$ we use the extrinsic curvature tensor definition $k_{ij}:=\pm( n^{\mu}\partial_{\mu}\gamma_{ij})/2$. Its trace $k:=\gamma^{ij}k_{ij}$ enters the boundary Lagrangian which is defined as $\mathcal{L}_{\partial M}:=\mathcal{L}_{\partial M}[\gamma_{ij}, k_{ij}]=-k+\mathcal{L}_{ct}$. When $(M, g_{\mu \nu})$ contains a radial coordinate $r\in[0,\infty)$, the boundary counter-term $\mathcal{L}_{ct}$ ensures that $-\beta F=-I^{E}$ is finite as $r\rightarrow \infty$. Consequently, considering $-I^{E}$ as a functional of $\gamma_{ij}$ allows for the derivation of the regularised Brown-York tensor $\bar{\Pi}_{ij}$ as
\begin{equation}
    \bar{\Pi}_{ij}:=\frac{2}{\sqrt{\gamma}}\frac{\delta(-I^{E})}{\delta\gamma^{ij}}=\frac{1}{8\pi}\bigg(k_{ij}-k\gamma_{ij} + \frac{\delta (-I^{E}_{ct})}{\delta \gamma^{ij}}\bigg) \ .
    \label{BY0}
\end{equation}
If the contraction of the $3^{\rm rd}$ term in (\ref{BY0}) with $\gamma^{ij}$ is proportional to $\mathcal{L}_{ct}$, then taking the trace of (\ref{BY0}) yields  $\bar{\Pi}/(d-2)=\mathcal{L}_{\partial M}/8\pi$. Therefore, the boundary integral $-I^{E}_{\partial M}$ in (\ref{I1}) is $\bar{\Pi}/2$ integrated over $(\partial M, \gamma_{ij})$.\

When $(M, g_{\mu \nu})$ is defined by the line element $ds^{2}_{M}=f(r)d\tau^{2}+f(r)^{-1}dr^{2}+r^{2}d\Omega_{d-2}$, we choose $(\partial M ,\gamma_{ij})$ to be a constant $r=r_{c}$ hypersurface such that $ds^{2}_{\partial M}=f(r_{c})d\tau^{2}+r_{c}^{2}d\Omega_{d-2}$ with volume element $\sqrt{\gamma}=\sqrt{f(r_{c}})\times\sqrt{\sigma}$, where $\sqrt{\sigma}$ is the area element on $S^{d-2}$. Accordingly, the boundary integral measure in (\ref{I1}) is $d^{d-1}x\sqrt{\gamma}=d\tau\sqrt{f(r_{c}})\ d^{d-2}x\sqrt{\sigma}$. and the boundary integral in (\ref{I1}) can be evaluated as $-I^{E}_{ \partial M}=\beta_{H}\sqrt{f(r_{c}}) ({\rm Area}(S^{d-2}(r_{c})(\bar{\Pi}/d-2))$. If the bulk integral $-I^{E}_{M}$ vanishes when evaluated on classical saddles, then we may make the identification $\beta F=-\beta_{H}\sqrt{f(r_{c}}) ({\rm Area}(S^{d-2}(r_{c}))(\bar{\Pi}/d-2))$ \cite{York, BrownYork}. Naturally, it follows that the relevant thermodynamic temperature on $(\partial M, \gamma_{ij})$ is the Tolman-Ehrenfest temperature $T_{TE}:=T_{H}/\sqrt{f(r_{c})}$ \cite{TE1, Sant1, Sant2, Sant3}, where $T_{H}$ is the global temperature $T_{H}:=f'(r_{H})/4\pi$ \cite{Hawk3}. As fixing $r=r_{c}$ fixes both $T_{TE}$ and the geometric volume of $(\partial M, \gamma_{ij})$, we can identify the intensive helmholtz free energy $f$ and its extensive counterpart $F$ on $(\partial M, \gamma_{ij})$ as
\begin{align}
    \begin{split}
        f=-\frac{\bar{\Pi}}{d-2} \rightarrow F={\rm Area}(S^{d-2}(r_{c}))\bigg(-\frac{\bar{\Pi}}{d-2}\bigg) \ .
        \label{FE11}
    \end{split}
\end{align}
Where we depart from the traditional quasi-local analysis is that the thermal system under consideration is  $(\partial M, \gamma_{ij})$ only, rather than $(\partial M, \gamma_{ij})$ as the end of an extended system including $(M,g_{\mu \nu})$ to the interior. Furthermore, the central premise of this analysis is that the following intensive Euler relation holds exclusively on $(\partial M, \gamma_{ij})$ \cite{Isichei}
\begin{equation}
f=\rho-Ts+\mu_{I}Q_{I} \ ,
\label{Euler}
\end{equation}
 where $\rho$ is energy density , $T$ is temperature and $\mu_{I}$ and $Q_{I}$ are chemical potentials and charges, the number of which $I$, is geometry dependent.  Granting this premise and matching (\ref{Euler}) with (\ref{FE11}) results in the thermodynamic potentials in the Euler relation on $(\partial M, \gamma_{ij})$ being entirely determined by the constituents of $-\bar{\Pi}/(d-2)$. Accordingly, we observe that from the individual angular components $\bar{\Pi}^{a}_{a}$, a quasi-local ``entropy" function $s$ in  (\ref{Euler}) can be identified. Again, we stress that identifying (\ref{FE11}) as $F_{QL}=E_{QL}-T_{TE}S$ necessarily informs one about thermal properties of $(M, g_{\mu \nu})$ in the interior. We avoid this in favour of the ansatz (\ref{Euler}) where $(\partial M, \gamma_{ij})$ alone is the thermal system.   \

We believe that this prescription holds for any static, spherically symmetric black hole geometry in arbitrary dimensions\footnote{As long as $\bar{\Pi}_{ij}$ is finite and non-vanishing throughout $(M, g_{\mu \nu})$}. Nevertheless, for the sake of clarity we will choose specific geometries in specific dimensions to elucidate various aspects of this general result.\

\textbf{\textit {Euclidean Schwarzschild black hole in d=4:}} When\footnote{ The event horizon is defined as $r_{0}=2M$ when $\hbar=c=G=1$} $f(r)=1-r_{0}/r$ the bulk Lagrangian takes the form $\mathcal{L}_{M}=R$. The bulk action $I^{E}_{M}$ vanishes in the saddle-point approximation since $R=0$. Consequently, the only contribution to the free energy is from $-I^{E}_{\partial M}$. The regularisation of $\mathcal{L}_{\partial M}$ is achieved via Euclidean reference subtraction in which $\mathcal{L}_{ct}=+k_{ET}$ where $k_{ET}$ is the extrinsic curvature trace of Euclidean thermal geometry. Using this regularisation scheme, the boundary Lagrangian is $\mathcal{L}_{\partial M}=k_{ET}-k:=[k]$. Subsequently, $\bar{\Pi}_{ij}$ can be derived from $-I^{E}$ according to (\ref{BY0}) \cite{Israel1, Israel2}
\begin{equation}
    \bar{\Pi}_{ij}=\frac{1}{8\pi}\bigg([k_{ij}]-[k]\gamma_{ij}\bigg) \ .
    \label{BY1}
\end{equation}
Taking the trace of (\ref{BY1}) reveals that $ \bar{\Pi}/2=[k]/8\pi=\mathcal{L}_{\partial M}$. Additional insight is gained by computing $\bar{\Pi}^{i}_{j}=\gamma^{ik}\bar{\Pi}_{jk}$, the individual constituents of the trace which are \cite{F0}
\begin{equation}
    \bar{\Pi}^{\tau}_{\tau}=\frac{-1}{4\pi r_{c}}\Bigg(1-\sqrt{1-\frac{r_{0}}{ r_{c}}}\Bigg) \ \ \& \ \ \bar{\Pi}^{\theta}_{\theta}=\frac{1}{8 \pi r_{c}}\Bigg(\frac{1-\frac{r_{0}}{2r_{c}}}{\sqrt{1-\frac{r_{0}}{r_{c}}}}-1\Bigg),
    \label{BY2}
\end{equation}
where the individual angular constituents $\bar{\Pi}^{\theta}_{\theta}$ and $\bar{\Pi}^{\phi}_{\phi}$ are degenerate \cite{York, BrownYork, BrownYork2}. On $(\partial M, \gamma_{ij})$ therefore, the intensive and extensive free energies take the relatively simple form 
\begin{align}
    \begin{split}
        &f=-\frac{\bar{\Pi}^{\tau}_{\tau}}{2}-\bar{\Pi}^{\theta}_{\theta} \rightarrow 
        F=4\pi r_{c}^{2}\Bigg(-\frac{\bar{\Pi}^{\tau}_{\tau}}{2}-\bar{\Pi}^{\theta}_{\theta}\Bigg) \ .
    \end{split}
    \label{FE12}
\end{align}
Keeping both (\ref{FE11}) and the ansatz (\ref{Euler}) in mind, we consider the we consider $(\partial M, \gamma_{ij})$ and the observer to be synonymous. Therefore, we reinterpret $-I^{E}_{\partial M}$ as the integral along the world-line of a static observer. According to (\ref{FE11}) and (\ref{Euler}), the observer $(\partial M, \gamma_{ij})$ sees a localised energy density $\rho=-\bar{\Pi}^{\tau}_{\tau}/2$. The angular components $\bar{\Pi}^{\theta}_{\theta}=\bar{\Pi}^{\phi}_{\phi}$ are usually interpreted as a transverse pressure $p_{T}$. Physically, $p_{T}$ is a surface gravity which counteracts the local gravitational acceleration $(\partial M, \gamma_{ij})$ experiences so that $(\partial M, \gamma_{ij})$ remains at $r=r_{c}$ \cite{F1}. The form of $\bar{\Pi}^{\theta}_{\theta}$ in (\ref{BY2}) shows that $p_{T}\rightarrow \infty$ as $r_{c}\rightarrow r_{0}$ while conversely, $p_{T}\rightarrow 0$ as $r_{c}\rightarrow \infty$. Normally, $p_{T}$ is considered as being thermodynamically conjugate to $4\pi r_{c}^{2}$, the area of $(\partial M, \gamma_{ij})$. Nonetheless, we espouse an alternative perspective of $p_{T}$ on $(\partial M, \gamma_{ij})$ which is as follows; recall that in gravitational thermodynamics, geometric concepts are reinterpreted as thermal quantities. Enforcing the regularity of euclideanised near horizon geometry allows for the reinterpretation of the horizon surface gravity (from infinity) $\kappa=1/(2r_{0})$ as the Hawking temperature $T_{H}:=f'(r_{H})/4\pi=1/(4\pi r_{0})$ \cite{Hawk3}. The temperature $T_{TE};=T_{H}/\sqrt{f(r_{c})}$ is simply the blue-shifted asymptotic surface gravity at finite $r_{c}$. Additionally, the event horizon area is reinterpreted as an entropy. As $p_{T}$ is a surface gravity acting over the geometric area of $(\partial M, \gamma_{ij})$, a completely reasonable question to ask is how $p_{T}$ is related to $T_{TE}$ and entropy. We observe that, to reconcile (\ref{Euler}) and (\ref{FE12}) when $T=T_{TE}$, it must follow that
\begin{equation}
    p_{T}=T_{TE}s \ ,
    \label{PTS}
\end{equation}
where $s$ is an intensive quasi-local entropy density localised to $(\partial M, \gamma_{ij})$ at $r=r_{c}$.  The divergence of $p_{T}$ at $r=r_{0}$ is now imputed to the divergence of $T_{TE}$ at the horizon $r_{0}$, while $s$ remains finite \cite{Isichei, F2}. According to (\ref{PTS}), the individual angular components $\bar{\Pi}^{\theta}_{\theta}$ and $\bar{\Pi}^{\phi}_{\phi}$ can be reinterpreted as the product of $T_{TE}=1/(4\pi r_{0}\sqrt{f(r_{c})})$ and the intensive quasi-local entropy on $(\partial M, \gamma_{ij})$
\begin{equation}
    s=\frac{r_{0}}{2r_{c}}\Bigg(1-\frac{r_{0}}{2r_{c}}-\sqrt{1-\frac{r_{0}}{r_{c}}}\Bigg)\ .
    \label{E1}
\end{equation}
This entropy density vanishes at spatial infinity and monotonically increases as $r_{c}$ decreases. On the event horizon $r_{c}=r_{0}$, $s$ takes its maximum value $s(r_{0})=1/4$ such that the extensive entropy $S=4\pi r_{c}^{2}s$ is the Bekenstein-Hawking entropy $S=\pi r_{0}^{2}$. As $s<1/4$ for $r_{c}>r_{0}$, then the extensive entropy must obey $S<\pi r_{c}^{2}$ for $r_{c}>r_{0}$, an explicit verification of the covariant entropy conjecture. Moreover, consider the aforementioned case of an ingoing radial null geodesic with $\theta<0$ beginning on $(\partial M, \gamma_{ij})$ at $r=r_{2}$ and ending at $(\partial M', \gamma_{ij}')$ at $r=r_{1}$. Using (\ref{E1}) the entropy of the system can be calculated as $S_{L}=4\pi r_{2}^{2}s(r_{2})-4\pi r_{1}^{2}s(r_{1})$. Since $s(r_{2})<1/4$ and $s(r_{1})<1/4$ for $r_{2},r_{1}>r_{0}$
, then it immediately follows that $S_{L}<\pi r_{2}^{2}-\pi r_{1}^{2}$ which is the entropy bound for this scenario \cite{Wald}. For the sake of consistency, we interpret $S_{L}$ as the entropy difference of the two disconnected boundaries $(\partial M, \gamma_{ij})$ and $(\partial M', \gamma_{ij}')$, rather than the entropy of the intervening geometry. Thus, the quasi-local entropy (\ref{E1}) allows for the explicit evaluation of the entropy of a thermal system consisting of least one boundary $(\partial M, \gamma_{ij})$ which obeys (\ref{Euler}) and (\ref{FE12}).\

\textbf{\textit {Euclidean Reissner-Nordstrom black hole in d=4:}} In this case $(M, g_{\mu \nu})$ is defined by $f(r)=(r-r_{+})(r-r_{-})/r^{2}$ where $r_{\pm}=r_{0}/2\pm \sqrt{r_{0}^{2}/4-Q^{2}}$ and $Q$ is the global black hole charge. The bulk Lagrangian is $\mathcal{L}_{M}=R-F_{\mu \nu}F^{\mu \nu}$ where $F_{\mu \nu}=\nabla_{\mu}A^{E}_{\nu}-\nabla_{\nu}A^{E}_{\mu}$ is the Maxwell tensor for the Euclideanised gauge potential $A^{E}_{\mu}dx^{\mu}:=iA_{\mu}dx^{\mu}=i(Q/r-\Phi)d\tau$ and $\Phi=Q/r_{+}$ is the potential on the horizon $r_{+}$ \cite{Hawk1}. The reference subtracted boundary Lagrangian takes the form $\mathcal{L}_{\partial M}=[k]$. Integrating $F_{\mu\nu}F^{\mu \nu}$ by parts such that $-F_{\mu\nu}F^{\mu \nu}=-2A^{E}_{i}n_{j}F^{ij}+2A^{E}_{\mu}\nabla_{\nu}F^{\mu \nu}$ leads to  $\mathcal{L}_{M}=R+2A^{E}_{\nu}\nabla_{\mu}F^{\mu \nu}$ and $\mathcal{L}_{\partial M}=[k]-2A^{E}_{i}n_{j}F^{ij}$. As in the Schwarzschild case, $-I^{E}_{M}$ vanishes in the saddle point approximation as $R=0$ and $\nabla_{\nu}F^{\mu \nu}=0$ hold classically so that the only contribution to $-\beta F$ is from $-I^{E}_{\partial M}$ \cite{Hawk1, Hawk4, BrownYork3, BrownYork4}.\\
As $\mathcal{L}_{ct}=+k_{ET}$ is chosen, then $\bar{\Pi}_{ij}$ is derived from $-I^{E}$ in the exact same manner as the Schwarzschild case shown in (\ref{BY1}). Therefore the regularised trace obeys the relation $\bar{\Pi}/2=[k]/8\pi$. Yet, due to the electromagnetic contribution to $\mathcal{L}_{\partial M}$, $f$ is not fully determined by $-\bar{\Pi}/2$. However, important thermodynamic information on $(\partial M, \gamma_{ij})$ may still be ascertained from the individual trace components $\bar{\Pi}^{i}_{i}$ which are \cite{F0}
\begin{align}
\begin{split}
    &\bar{\Pi}^{\tau}_{\tau}=\frac{-1}{4\pi r_{c}}\Bigg(1-\frac{\sqrt{(r_{c}-r_{+})(r_{c}-r_{-})}}{r_{c}}\Bigg) \ \&  \\
    &\bar{\Pi}^{\theta}_{\theta}=\frac{1}{8 \pi r_{c}}\Bigg(\frac{(r_{c}-r_{+})+(r_{c}-r_{-})}{2\sqrt{(r_{c}-r_{+})(r_{c}-r_{-})}}-1\Bigg) ,
\end{split}
\label{BYRN}
\end{align}
where $\bar{\Pi}^{\theta}_{\theta}=\bar{\Pi}^{\phi}_{\phi}$. Evaluating the the boundary term $-2A^{E}_{i}n_{j}F^{ij}$ on $(\partial M, \gamma_{ij})$ and taking its sum with $-\bar{\Pi}/2$ yields the total free energy on the surface
\begin{align}
    \begin{split}
        &f=\frac{-\bar{\Pi}^{\tau}_{\tau}}{2}-\bar{\Pi}^{\theta}_{\theta}+\frac{iA^{E}_{\tau}Q}{8\pi r_{c}\sqrt{(r_{c}-r_{+})(r_{c}-r_{-})}}\ ,
    \end{split}
    \label{FE13}
\end{align}
where, according to (\ref{Euler}), we interpret $\rho=-\bar{\Pi}^{\tau}_{\tau}/2$, while the $3^{\rm rd}$ term in (\ref{FE13}) is identified with the chemical potential charge product $\mu_{I}Q_{I}$ where $I=1$. If the chemical potential is identified as $\mu_{1}=iA_{\tau}^{E}/\sqrt{f(r_{c})}$ \cite{BrownYork3, Huang}, then the charge density on $(\partial M, \gamma_{ij})$ must be identified as $Q_{1}=Q/8\pi r_{c}^{2}$. Once again, the conventional interpretation of $\bar{\Pi}^{\theta}_{\theta}$ is a transverse pressure $p_{T}$ which acts as a surface gravity along $(\partial M, \gamma_{ij})$ and is conjugate to the area of $(\partial M, \gamma_{ij})$. As in the Schwarzschild case, this standard result will be subject to a non-standard thermal reinterpretation which, in this case, is region dependent.\

\underline{$r_{-}<r_{+}\leq r_{c}$}: In this region $f(r)>0$ such that the Euclidean time coordinate $\tau$ is space-like. Consequently, the temperature of relevance is $T_{H(+)}:=f'(r_{\pm})/4\pi$ for this $f(r)$ is $T_{H(+)}=(r_{+}-r_{-})/4\pi r^{2}_{+}>0$ \cite{Hawk1} . Naturally, $T_{H(+)}$ as the positive temperature of the outer horizon $r_{+}$ is considered as the global temperature of $(M,g_{\mu \nu})$ when  $(\partial M, \gamma_{ij})$ is at spatial infinity. Nevertheless, when $(\partial M, \gamma_{ij})$ is at a finite $r_{c}$, one may still consider the local temperature $T_{TE (+)}=T_{H(+)}/\sqrt{f(r_{c})}$ \cite{BrownYork3, Huang}. As $r_{-}<r_{+}\leq r_{c}$, it follows from (\ref{BYRN}) that $\bar{\Pi}^{\theta}_{\theta}>0$ in this region. In identical fashion to the Schwarzschild case, equating (\ref{FE13}) and (\ref{Euler}) leads to the identification (\ref{PTS}). Therefore, $\bar{\Pi}^{\theta}_{\theta}=p_{T}$ in (\ref{BYRN}) may be reinterpreted as the product of  $T_{TE(+)}$ and an intensive quasi-local entropy density on $(\partial M, \gamma_{ij})$ for Reissner-Nordstrom geometry given by
\begin{align}
    \begin{split}
        s_{RN(+)}=\frac{r_{+}^{2}}{4r_{c}^{2}}\Bigg(\frac{(r_{c}-r_{+})+(r_{c}-r_{-})}{r_{+}-r_{-}}\Bigg)\\-\frac{r_{+}^{2}}{2r_{c}^{2}}\Bigg(\frac{\sqrt{(r_{c}-r_{+})(r_{c}-r_{-})}}{r_{+}-r_{-}}\Bigg) \ ,
        \label{E2}
    \end{split}
\end{align}
where $s_{RN(+)}$ is of relevance when $r_{+}\leq r_{c}<\infty$. The extensive equivalent of this entropy density on $(\partial M, \gamma_{ij})$ is $S=4\pi r_{c}^{2}s_{RN(+)}$. When $(\partial M, \gamma_{ij})$ is at spatial infinty $s_{RN(+)}$ vanishes. As $r_{c}$ decreases, $s_{RN(+)}$ monotonically increases until $(\partial M, \gamma_{ij})$ is on $r_{+}$. On $r_{+}$, $s_{RN(+)}$ takes its maximum value $s_{RN}(r_{+})=1/4$ so that the extensive entropy is the Bekenstein-Hawking entropy $S=\pi r_{+}^{2}$. Furthermore, it's apparent that $S<\pi r_{c}^{2}$ for $r_{c}>r_{+}$, which verifies the covariant entropy conjecture. Additionally, the entropy difference of the surfaces $(\partial M, \gamma_{ij})$ at $r_{c}=r_{2}$ and $(\partial M', \gamma_{ij}')$ at $r_{c}=r_{1}$ where $r_{1}<r_{2}$ is $S_{L}=4\pi r_{2}^{2}s_{RN(+)}(r_{2})-4\pi r_{2}^{2}s_{RN(+)}(r_{1})$ where once again, $S_{L}<\pi r_{2}^{2}-\pi r_{1}^{2}$ for $r_{1},r_{2}>r_{+}$ \cite{Wald}.\

\underline{$r_{-}<r_{c}<r_{+}$:} In this region $f(r)<0$ so that $\tau$ is no longer space-like. This, in turn, leads to the intensive free energy in (\ref{FE13}) being complex. As a result, thermal analysis is not possible in this region.\

\underline{$r_{c} \leq r_{-}<r_{+}$:} Here, $f(r)>0$, so that $\tau$ regains its space-like nature. Accordingly, the intensive free energy (\ref{FE13}) is once again real. However, in this region $\bar{\Pi}^{\theta}_{\theta}=p_{T}<0$ so that $p_{T}$ should now be considered as a surface tension acting over $(\partial M, \gamma_{ij})$ to keep $r_{c}$ fixed. Relatedly, the temperatures of relevance are $T_{H(-)}=(r_{-}-r_{+})/4\pi r_{-}^{2}<0$ and its associated local temperature $T_{TE}=T_{H(-)}/ \sqrt{f(r_{c}})<0$. Using (\ref{PTS}), we interpret the surface tension $p_{T}$ as the product of the negative $T_{TE(-)}$ and the positive entropy density 
\begin{align}
    \begin{split}
        s_{RN(-)}=\frac{r_{-}^{2}}{4r_{c}^{2}}\Bigg(\frac{(r_{c}-r_{+})+(r_{c}-r_{-})}{r_{-}-r_{+}}\Bigg)\\-\frac{r_{-}^{2}}{2r_{c}^{2}}\Bigg(\frac{\sqrt{(r_{c}-r_{+})(r_{c}-r_{-})}}{r_{-}-r_{+}}\Bigg) \ .
        \label{E3}
    \end{split}
\end{align}
This region exhibits rather unwonted thermal behaviour as $T_{TE(-)}$ ranges from $T_{TE(-)}=-\infty$ at $r_{c}=r_{-}$ to $T_{TE}\rightarrow0$ as $r_{c}\rightarrow0$. It follows that the zero temperature singularity $r=0$ is the ``hottest" part of this region classically. Naturally, entropy should increase going from ``cold" to ``hot" which is true of (\ref{E3}). However, in contrast to (\ref{E1}) and (\ref{E2}), $s_{RN(-)}$ takes a minimum value of $s_{RN(-)}(r_{-})=1/4$ and increases without bound with $s_{RN(-)}\rightarrow \infty$ as $r_{c}\rightarrow 0$. As the extensive entropy obeys $S=4\pi r_{c}^{2}s_{RN(-)}>\pi r_{c}^{2}$ for $r_{c}<r_{-}$, the covariant entropy conjecture is violated in this region. We do not consider this as an invalidation of the conjecture, but rather, a logical consequence of ascribing negative temperature to this region.

\textbf{\textit{Euclidean ${\rm \bf AdS}$ Black Hole in ${\bf d=5}$:}}  $(M, g_{\mu \nu})$ is defined by $f(r)=1-(r_{0}/r)^{2}-(r/l)^{2}$ where $l$ is the AdS radius. Consequently, the bulk Lagrangian takes the form $\mathcal{L}_{M}=R_{(M)}+12/l^{2}$.  In contrast to the previous cases, dimension dependent covariant counter-term subtraction is utilised with $\mathcal{L}_{ct}=-3/l-R_{(\partial M)}l/4$, so that $\mathcal{L}_{\partial M}=-k-3/l- R_{(\partial M)}l/4$, where $R_{(M)}$ and $R_{(\partial M)}$ are the $5d$ bulk and $4d$  boundary Ricci scalars respectively \cite{Vijay}. This choice of $\mathcal{L}_{ct}$ ensures that $-I^{E}$ is finite as $r_{c}\rightarrow \infty$. However, at finite $r_{c}$ we don't necessarily need to consider $\mathcal{L}_{ct}$. Nonetheless, we will include $\mathcal{L}_{ct}$ for illustrative purposes. As such, the regularised Brown-York tensor $\bar{\Pi}_{ij}$ in $5d$ is \cite{Vijay}
\begin{equation}
    \bar{\Pi}_{ij}=\frac{1}{8\pi }\bigg(k_{ij}-k\gamma_{ij}-\frac{3}{l}\gamma_{ij}+\frac{l}{2}G_{ij(\partial M)}\bigg),
    \label{BYADS}
\end{equation}
where $G_{ij (\partial M)}=R_{ij (\partial M)}-\gamma_{ij}R_{(\partial M)}/2$ is the Einstein tensor on $(\partial M, \gamma_{ij})$ and $k_{ij}=-( n^{\mu}\partial_{\mu}\gamma_{ij})/2$ is the extrinsic curvature along the inward pointing normal. The constitutents of the trace are
\begin{align}
    \begin{split}
        &\bar{\Pi}^{\tau}_{\tau}=\frac{-1}{8\pi}\Bigg(\frac{3}{l}+\frac{3l}{2r_{c}^{2}}-\frac{3}{r_c}\sqrt{1-\bigg(\frac{r_{0}}{r_{c}}\bigg)^{2}+\bigg(\frac{r_{c}}{l}\bigg)^{2}}\Bigg) \ \& \\
        &\bar{\Pi}_{\theta}^{\theta}=\frac{1}{8\pi}\Bigg(\Bigg(\frac{r_{0}^{2}}{r_{c}^{3}}+\frac{r_{c}}{l^{2}}\Bigg)\sqrt{1-\bigg(\frac{r_{0}}{r_{c}}\bigg)^{2}+\bigg(\frac{r_{c}}{l}\bigg)^{2}}^{\ -1}\\& \quad \quad +\frac{2}{r_{c}}\sqrt{1-\bigg(\frac{r_{0}}{r_{c}}\bigg)^{2}+\bigg(\frac{r_{c}}{l}\bigg)^{2}}-\frac{3}{l}-\frac{l}{2r_{c}^{2}}\Bigg) \quad,
        \label{BYADS2}
    \end{split}
\end{align}
where $\bar{\Pi}_{\theta}^{\theta}=\bar{\Pi}_{\phi}^{\phi}=\bar{\Pi}_{\theta}^{\theta}$ on $S^{3}$. Contrary to the previous two cases in which Euclidean reference subtraction was used, by taking the trace of (\ref{BYADS}), we notice that $\bar{\Pi}/3\neq \mathcal{L}_{\partial M}/(8\pi)$. This is entirely due to $\mathcal{L}_{ct}$ being composed of terms covariant on $(\partial M, \gamma_{ij})$ rather than extrinsic curvature terms. A simple remedy for this is to still consider the intensive free energy on $(\partial M, \gamma_{ij})$ as $f=-\mathcal{L}_{\partial M}/(8\pi)$ which is expressed in terms of trace constitutents in (\ref{BYADS2}) as
\begin{equation}
    f=\frac{-\bar{\Pi}^{\tau}_{\tau}}{3}-\bar{\Pi}^{\theta}_{\theta}+f_{\rm cov}, \ {\rm where }, \ f_{\rm cov}=\frac{l}{12}-\frac{1}{l} \ .
\end{equation}
Once again according to (\ref{Euler}), we identify $\rho=-\bar{\Pi}^{\tau}_{\tau}/3$ and $\bar{\Pi}^{\theta}_{\theta}=\bar{\Pi}^{\phi}_{\phi}=\bar{\Pi}^{\psi}_{\psi}=p_{T}$ on $(\partial M, \gamma_{ij})$. To apply the decomposition in (\ref{PTS}), we use the fact that the global temperature $T_{H}=f'(r_{H})/4\pi= (r_{0}^{2}/r_{H}^{3}+r_{H}/l^{2})/2\pi$ so that $T_{TE}=T_{H}/\sqrt{f(r_{c})}$ \cite{Brown}. Hence, one may derive the intensive quasi-local entropy density for ${\rm AdS_{5}}$ geometry from $\bar{\Pi}^{\theta}_{\theta}$ in (\ref{BYADS2}) using (\ref{PTS}) which yields
\begin{align}
    \begin{split}
        s_{AdS}=&\frac{1}{4}\bigg(\frac{r_{0}^{2}}{r_{H}^{3}}+\frac{r_{H}}{l^{2}}\bigg)^{-1}\Bigg(\bigg(\frac{r_{0}^{2}}{r_{c}^{3}}+\frac{r_{c}}{l^{2}}\bigg)\\&+\frac{2}{r_{c}}\bigg(1-\bigg(\frac{r_{0}}{r_{c}}\bigg)^{2}+\bigg(\frac{r_{c}}{l}\bigg)^{2}\bigg)\\
        &-\bigg(\frac{3}{l}+\frac{l}{2r_{c}^{2}}\bigg)\sqrt{1-\bigg(\frac{r_{0}}{r_{c}}\bigg)^{2}+\bigg(\frac{r_{c}}{l}\bigg)^{2}}\Bigg) \ .
    \end{split}
\end{align}
Once again, we see that $s_{AdS}(r_{H})=1/4$ such that $S=2\pi ^{2}r_{H}^{3}s_{AdS}(r_{H})$ is the Bekenstein-Hawking entropy for $AdS_{5}$, with $S<\pi^{2}r_{H}^{3}/2$ for $r>r_{H}$ verifying the covariant entropy conjecture. In contrast to the previous cases, $s_{AdS}$ asymptotes to a constant value at spatial infinity since $p_{T}$ is non-zero globally \cite{Vijay}. However, in this regime we no longer consider (\ref{Euler}) and $(\ref{PTS})$ as valid.

\newpage
\textit{Acknowledgements:} We would like to thank Will Chan, Michael Fine, Joao Magueijo, Jaeha Park and Andrew Svesko for informative discussions. The original draft of this letter was written as an essay for the ``2026 Awards for Essays on Gravitation" for which it received an honourable mention. We are also indebted to the anonymous reviewer at IJMPD for many helpful comments. RI is supported by the Institute of Physics through the Bell Burnell Graduate Scholarship Fund. No artificial intelligence was in the derivation of results in this letter, or preparation of the manuscript.

\clearpage

\appendix


\begin{thebibliography}{99}

\bibitem{Hawk1}
G.~W.~Gibbons and S.~W.~Hawking,
Phys.\ Rev.\ D \textbf{15}, 2752 (1977),
doi.org/10.1103/PhysRevD.15.2752.

\bibitem{Hawk2}
J.~M.~Bardeen, B.~Carter and S.~W.~Hawking,
Commun.\ Math.\ Phys.\ \textbf{31}, 161 (1973),
doi.org/10.1007/BF01645742.

\bibitem{Hawk3}
S.~W.~Hawking,
Commun.\ Math.\ Phys.\ \textbf{43}, 199 (1975),
doi:10.1007/BF02345020.

\bibitem{Bek1}
J.~D.~Bekenstein,
Lett.\ Nuovo.\ Cimento.\ \textbf{4}, 737 (1972),
doi.org/10.1007/BF02757029.

\bibitem{Bek2}
J.~D.~Bekenstein,
Phys.\ Rev.\ D \textbf{7}, 2333 (1973),
doi:10.1103/PhysRevD.7.2333.

\bibitem{Bek3}
J.~D.~Bekenstein,
Phys.\ Rev.\ D \textbf{9}, 3292 (1974),
doi.org/10.1103/PhysRevD.9.3292.

\bibitem{York}
J.~W.~York,
Phys.\ Rev.\ D \textbf{33}, 2092 (1986),
doi.org/10.1103/PhysRevD.33.2092.

\bibitem{BrownYork}
J.~D.~Brown and J.~W.~York,
Phys.\ Rev.\ D \textbf{47}, 1407 (1993),
doi.org/10.1103/PhysRevD.47.1407.

\bibitem{BrownYork2}
J.~D.~Brown and J.~W.~York,
[arXiv:gr-qc/9405024 [gr-qc]].

\bibitem{TE1}
R.~Tolman and P.~Ehrenfest,
Phys.\ Rev.\  \textbf{36}, 1791 (1930),
doi.org/10.1103/PhysRev.36.1791.

\bibitem{Sant1}
J.~Santiago and M.~Visser,
Eur.\ J.\ Phys \textbf{40}, 025604 (2019),
doi.org/10.1088/1361-6404/aaff1c,
[arXiv:1803.04106 [gr-qc]].

\bibitem{Sant2}
J.~Santiago and M.~Visser,
Int.\ J.\ Mod.\ Phys.\ D \textbf{14}, 1846001, (2018),
doi.org/10.1142/S021827181846001X,
[arXiv:1805.05583 [gr-qc]].

\bibitem{Sant3}
J.~Santiago and M.~Visser,
Phys.\ Rev.\ D \textbf{98}, 064001 (2018),
doi.org/10.1103/PhysRevD.98.064001,
[arXiv:1807.02915 [gr-qc]].

\bibitem{Bek4}
J.~D.~Bekenstein,
Phys.\ Rev.\ D \textbf{23}, 287 (1981),
doi.org/10.1103/PhysRevD.23.287.

\bibitem{Bek5}
M.~Schiffer and J.~D.~Bekenstein,
Phys.\ Rev.\ D \textbf{39}, 1109 (1989),
doi.org/10.1103/PhysRevD.39.1109.

\bibitem{Suss1}
L.~Susskind,
J.\ Math.\ Phys \textbf{36}, 6377 (1995),
doi.org/10.1063/1.531249,
[arXiv:hep-th/9409089[hep-th]].

\bibitem{Suss2}
W.~Fischler and L.~Susskind,
[arXiv:hep-th/9806039 [gr-qc]].

\bibitem{thooft}
G.~'t Hooft,
[arXiv:gr-qc/9310026[gr-qc]].

\bibitem{Bousso}
R.~Bousso,
JHEP \textbf{07}, (1999) 004,
doi.org/10.1088/1126-6708/1999/07/004,
[arXiv:hep-th/9905177 [hep-th]].

\bibitem{Wald}
E.~E.~Flanagan, D.~Marolf and R.~M.~Wald,
Phys.\ Rev.\ D \textbf{62}, 084035 (2000),
doi.org/10.1103/PhysRevD.62.084035,
[ arXiv:hep-th/9908070v4[hep-th]].

\bibitem{F-1}
Of course, $S=A_{H}/4<A_{B}/4$ for the single boundary case and $S_{L}=0<(A_{B}-A_{B'})/4$ for the two boundary case. However, the tension persists if space-time geometry outside the event horizon does have inherent entropy which is the stance we take.

\bibitem{Isichei}
R.~Isichei, J.~Magueijo,
Phys.\ Rev.\ D \textbf{112}, L101502 (2025),
doi.org/10.1103/zzg7-vjb8,
[arXiv:2507.03469 [gr-qc]].


\bibitem{Israel1}
W.~Israel,
Nuovo.\ Cimento.\ B \textbf{44}, 1 (1966),
doi.org/10.1007/BF02710419.

\bibitem{Israel2}
W.~Israel,
Nuovo.\ Cimento.\ B \textbf{48}, 463 (1967),
doi.org/10.1007/BF02712210.

\bibitem{F0}
Using the formula for extrinsic curvature along an inward pointing normal in static geometry $k_{ij}=-(n^{\mu}\partial_{\mu}\gamma_{ij})/2$.

\bibitem{F1}
$\delta$ function localised tensors such as (\ref{BY1}) can be interpreted as a perfect fluid of the form $\bar{\Pi}_{ij}=-\rho  u_{i}u_{j}+p_{T}\sigma_{ij}$ where $u=u^{i}\partial_{i}$ is a Euclidean-time like vector and $\sigma_{ij}=\sigma_{ab}$ is the metric on $S^{2}$. When $p_{T}=0$, $\bar{\Pi}_{ij}=-\rho u_{i}u_{j}$ is interpreted as ``pressureless dust"  in free-fall within the background geometries. In this case, the worldine of $(\partial M, \gamma_{ij})$ and the shell of matter no longer coincide. Therefore an additional term needs to be added to (\ref{I1}) as discussed in \cite{Vijay01, Vijay02}.



\bibitem{Vijay01}
V.~Balasubramanian, A.~Lawrence, J.~M.~Magan, M.~Sasieta,
Phys.\ Rev.\ X \textbf{14}, 011024 (2024),
doi.org/10.1103/PhysRevX.14.011024,
[arXiv:2212.02447 [hep-th]].

\bibitem{Vijay02}
V.~Balasubramanian, A.~Lawrence, J.~M.~Magan, M.~Sasieta,
Phys.\ Rev.\ Lett. \textbf{132}, 141501 (2024),
doi.org/10.1103/PhysRevLett.132.141501,
[arXiv:2212.08623 [hep-th]].

\bibitem{F2}
To restore fundamental constants to (\ref{PTS}) we first note that the prefactor in (\ref{BY0}) should be $1/(8\pi G)$. Furthermore, extrinsic curvature has dimensions $[k_{ij}]=L$, and $[k]=L^{-1}$. Consequently, when Newton's constant is restored, $\bar{\Pi}_{\theta \theta}$ has dimensions $[\bar{\Pi}_{\theta\theta}]=ML^{-2}T^{2}$ (extensive), while $[\bar{\Pi}^{\theta}_{\theta}]=ML^{-4}T^{2}$ (intensive). As $[T_{TE}]=K$ and intensive thermodynamic entropy (per unit area) has dimensions $[s]=MT^{-2}K^{-1}$, it follows that a constant of dimensions $L^{-4}T^{4}$ must be added to the RHS of (\ref{PTS}) for it to be valid dimensionally. We take this constant to be $1/c^{4}$ such that $p_{T}=T_{TE}s/c^{4}$ when fundamental constants are restored.


\bibitem{Hawk4}
S.~W.~Hawking and S.~F.~Ross,
Phys.\ Rev.\ D \textbf{52}, 5865 (1995),
doi.org/10.1103/PhysRevD.52.5865,
[arXiv:hep-th/9504019 [hep-th]].

\bibitem{BrownYork3}
J.~D.~Brown, E.~A.~Martinez and J.~W.~York,
Phys.\ Rev.\ Lett.  \textbf{66}, 2281 (1991)
doi.org/10.1103/PhysRevLett.66.2281.

\bibitem{BrownYork4}
H.~W.~Braden, J.~D.~Brown,, B.~F.~Whiting and J.~W.~York,
Phys.\ Rev.\ D \textbf{42}, 3376 (1990),
doi.org/10.1103/PhysRevD.42.3376.

\bibitem{Huang}
B.~H.~Huang, L.~Zhao,
[arXiv:2605.07236 [gr-qc]].

\bibitem{Vijay}
P.~Kraus, V.~Balasubramanian,
Commun.\ Math.\ Phys.\ \textbf{208}, 413 (1999),
doi.org/10.1007/s002200050764,
[arXiv:hep-th/9902121 [hep-th]].

\bibitem{Brown}
J.~D.~Brown, J.~Creighton and R.~.B~Mann,
Phys.\ Rev.\ D \textbf{50}, 6394 (1994),
doi.org/10.1103/PhysRevD.50.6394,
[arXiv:gr-qc/9405007 [gr-qc]]








\end{thebibliography}
\end{document}